\documentclass[aps,letterpaper,twocolumn,prl,amsmath]{revtex4}
\usepackage{amsxtra}
\usepackage{graphicx}

\DeclareMathOperator{\Tr}{Tr}
\begin{document}
\title{Does the wavefunction describe individual systems?}
\author{Antonio Di Lorenzo}

\affiliation{Department of Physics and Astronomy, Stony Brook
University, SUNY, Stony Brook, NY 11794-3800}
\begin{abstract}
We analyze the issue of the interpretation of the wavefunction, 
namely whether 
it should be interpreted as describing individual systems or ensembles 
of identically prepared systems. 
We propose an experiment which can decide the issue, 
based on the simultaneous measurement of the same observable with 
different detectors, 
and we discuss the theoretical implications of the possible experimental 
outcomes.
\end{abstract}
\maketitle
The question enunciated in the title is a central issue 
for the interpretation of Quantum Mechanics. 
It has a long history: 
in 1935, Einstein, Podolsky, and Rosen \cite{Einstein1935} 
demonstrated that Quantum Mechanics 
is an incomplete theory, i.e. that individual systems do possess 
objective properties which the wave-function does not account for. 
In order to agree with their conclusions, however, one must accept 
their definition of objectivity, which is 
\begin{quote}
If, without in any way disturbing a system, 
we can predict with certainty (i.e., with probability equal to unity) 
the value of a physical quantity, 
then there exists an element of reality corresponding to that quantity.
\end{quote}
In this paper, we shall introduce a 
weaker definition of objectivity, which we hope may be widely accepted:
\begin{quote}
If, by simultaneously measuring the same physical quantity of a system 
by means of several detectors, the outcomes of the latter agree, 
then the value of that physical quantity is an objective property of 
the system. 
\end{quote}
The ensemble interpretation has been held by a minority of the physics 
community, a minority that is now growing in size, at least in our perception. 
The dichotomy between the wavefunction describing ensembles or individual 
systems is believed to be a metaphysical issue, not susceptible of 
experimental resolution. In this note we propose an ideal experiment, 
leaning on the above definition of objectivity, 
which may provide the answer. 

First, we explain why the issue at hand is of great interest. 
Let us assume, for the moment being, that the wavefunction describes 
individual systems. 
Let us consider the paradigmatic example of an entangled system: 
a system composed of two spin 1/2 particles forming a singlet. 
When the two particles are 
far apart, and the spin of either of them is measured 
along an axis $\mathbf{n}$, 
giving an outcome $\pm \hbar/2$, then the wavefunction of the other 
particle instantaneously collapses into the state 
$|\mp\mathbf{n} \hbar/2\rangle$. 
Since we are assuming that the wavefunction 
describes individual systems, and thus possesses a physical reality, 
this is an instance of non-locality. 
From this point of view, Quantum Mechanics is a mysterious and surprising 
theory. 

On the other hand, if one interprets the wavefunction as describing ensembles, 
the collapse of the second particles' state means that, if one 
measures the spin of the second particle along a direction $\mathbf{n}'$,  
repeating
the measurement many times, the frequency with which the outcomes 
of the two measurements agree tends to 
$P_{ag}=(1+\mathbf{n}\cdot\mathbf{n}')/2$. 
Much of the surprise fades away, since 
the wavefunction appears now not as an objective property, but as a 
computational tool. 
The mystery, however, remains,   
since Bell has proved that the probability above cannot be reproduced by 
a local hidden variable theory (hidden variable theories  
attempt to describe individual systems by adding parameters 
having a stochastic distribution that is adjustable in order to match the 
predictions of Quantum Mechanics). Therefore, if one holds the view that 
Quantum Mechanics is incomplete, he must accept that a complete theory 
compatible with Quantum Mechanics must be nonlocal. 

The usual discussion of measurement involves a single detector measuring at 
different times an ensemble of identically prepared systems. 
We consider instead an ensemble of detectors simultaneously 
measuring the same individual system. 
For illustrative purposes, we consider a two-level system. 
Let $|0\rangle$ and $|1\rangle$ be the eigenstates of the measured quantity. 
Let there be $N$ detectors coupled to the system, 
which is prepared in the state 
\begin{equation}\label{cohstate}
c_0|0\rangle+c_1|1\rangle.
\end{equation} 
The question we ask is: 
What will be the output of the detectors in every single trial? 
We envision two possible scenarios compatible with Quantum Mechanics. 
In the first scenario, all the detectors give the same output, 
either $0$ or $1$. Upon repeating the measurement $M$ times on identically 
prepared systems, 
the detectors will all indicate $0$ 
a number of times $M_0\sim |c_0|^2 M$, and $1$ a number of times 
$M_1=M-M_0\sim |c_1|^2 M$ (the approximate equality accounts 
for an expected spread of $\sqrt{M}$ of the binomial distribution). 
In the second scenario, in each individual trial a fraction 
$N_0\sim |c_0|^2 N$ of the detectors will indicate $0$, 
and the remaining $N_1=N-N_0\sim |c_1|^2 N$ will indicate $1$. 
More precisely, the probability that $N_0$ detectors give 0 is 
\begin{equation}\label{binom}
P(N_0) = \binom{N}{N_0} |c_0|^{2 N_0} |c_1|^{2(N-N_0)}.
\end{equation}
This probability can be recovered by repeating the measurement several times. 
There is also a third scenario that we shall not consider: 
in each trial, the detectors will give different outcomes, 
however with a probability $P_3(N_0)$ 
different from $P(N_0)$ of Eq.\ \eqref{binom}, 
having still the property that $\sum N_0 P_3(N_0) = |c_0|^2 N$. 
We could not figure out how this last scenario could be make compatible 
with Quantum Mechanics, thus we shall leave it out of our discussion.

If the experiment reveals that the first case occurs, we will have 
evidence that the wavefunction describes ensembles and not individual system. 
We would conclude that an individual system has at each instant of time 
a well defined value of the observed variable, and that the superposition 
$c_0|0\rangle+c_1|1\rangle$ actually describes the preparation 
procedure. 
If the second case occurs, we will have evidence that the wavefunction 
does describe individual systems. 
 
Coupling a large number of detectors to a quantum system is a difficult 
task. We notice that it is sufficient to use two detectors only. 
If, upon repeating the measurement $M$ times, the detectors 
will give different outcomes a number of times $m\sim 2|c_0 c_1|^2 M$, 
we fall back into the second scenario, otherwise, if they always agree 
in a sufficiently large number of trials $M\gg 1/|c_0 c_1|^2$, 
the first scenario is confirmed. 
The issue of two simultaneous measurements was first discussed in 
Ref.~\onlinecite{Arthurs1965}, which discusses detection of complementary 
variables. More recently, Jordan and B\"uttiker\cite{Jordan2005} 
have considered two detectors simultaneously measuring (weakly)
the same quantity, and found a large value for the 
cross-correlations. 

We now proceed to predict the answer by applying Quantum Mechanics 
to the system and detectors.  
The procedure is the following: 
we let system and detectors interact, 
and calculate the time-evolution of the density matrix describing 
system and detectors.  
Then we trace out the system's degrees 
of freedom, and we consider the reduced density matrix of the detectors.  

As a toy model for measurement, we 
take the detectors to be harmonic oscillators, with 
Hamiltonian 
\begin{align}
H_{det}=&\ H_A^{(0)} + H_B^{(0)} \;\\
H_a^{(0)} =&\  
\frac{\Hat{p}_a^2}{2m_a} +
 \frac{1}{2}m_a\omega_a^2 \Hat{x}_a^2.
\end{align}
We take the interaction Hamiltonian to be 
\begin{equation}\label{int}
H_{int} =  
-\Hat{\sigma} \left(\lambda_A \Hat{x}_A + \lambda_B \hat{x}_B\right), 
\end{equation} 
where $\Hat{\sigma}$ is the measured observable, taking values $0,1$. 
We assume that $\Hat{\sigma}$ is conserved during the measurement. 
We also assume that initially the detectors are at equilibrium 
$\rho_{det}(0)=\exp{(-\beta H_{det})}/\Tr{\exp{(-\beta H_{det})}}$, 
their position being 
centered at $x_A=0$, $x_B=0$, with fluctuations, at temperatures 
higher than $\hbar\omega_a$,  
\begin{equation}
\Delta x_a^2= \frac{1}{\beta m_a\omega_a^2} \gg {\Delta_q} x_a^2 
:= \frac{\hbar}{m_a\omega_a},
\end{equation} 
where $\Delta_q x_a$ are the quantum fluctuations. 
If the initial density matrix of the system is, in the basis of the 
eigenstates of $\Hat{\sigma}$: 
\begin{equation}
\begin{pmatrix}
|c_0|^2&c_0c_1^*\\
c_0^*c_1&|c_1|^2
\end{pmatrix},
\end{equation} 
with $|c_0^2|+|c_1|^2=1$, 
then the reduced density matrix of the detectors is 
\begin{equation}\label{detev}
\rho_{det}(\tau) = |c_0|^2 \rho_{det}(0) + 
|c_1|^2 U_A U_B \rho_{det}(0) U_B^\dagger U_A^\dagger,
\end{equation}
where $U_a$ is the time evolution operator due to the Hamiltonian 
\[H_a^{(1)} = H_a+\lambda_a \Hat{x}_a  = 
\frac{\Hat{p}_a^2}{2m_a}+
\frac{1}{2}m_a\omega_a^2 (\Hat{x}_a-X_a)^2+const,\]
with \[X_a=\lambda_a/m_a\omega_a^2\] the equilibrium position 
corresponding to the outcome $\sigma=1$. 
If we include the influence of the environment in the time evolution,  
assuming that 
the measurement time is much larger than the relaxation time, 
we have that 
\begin{equation}\label{detev2}
\rho_{det}(\tau) = |c_0|^2 \rho_{A|0}\otimes \rho_{B|0}+ 
|c_1|^2 \rho_{A|1} \otimes \rho_{B|1}
+\mathcal{O}(e^{-\gamma\tau}),
\end{equation}
where $\gamma$ is the relaxation rate, 
and $\rho_{a|\sigma}= 
\exp{(-\beta H_a^{(\sigma)})}/\Tr{\exp{(-\beta H_a^{(\sigma)})}}$ 
is the density matrix of detector $a$ indicating  
the outcome $\sigma$. 
In order for the outcomes to be distinguishable from thermal noise 
we have to require that 
$X_a\gg \Delta x_a$, which ensures that the outcomes are also 
macroscopically distinguishable. 

Thus, when two detectors are 
simultaneously measuring the same two-valued observable, both will 
indicate $0$ with probability $|c_0|^2$ or $1$ with 
probability $|c_1|^2$. The extension to observables having 
more than two values is trivial. We point out that the result is independent 
of the detection model. In order to have the detectors' outcomes disagree 
with probability $2|c_0c_1|^2$, the reduced detector density matrix 
should read: 
\begin{align}\nonumber
\rho_{det} =&\ |c_0|^4 \rho_{A|0} \otimes \rho_{B|0} + 
|c_1|^4 \rho_{A|1} \otimes \rho_{B|1} \\ 
\label{eq:imp} &\ 
+|c_0 c_1|^2 \left(\rho_{A|0} \otimes \rho_{B|1}+ 
\rho_{A|1} \otimes \rho_{B|0}\right). 
\end{align}
Eq.~\eqref{eq:imp} violates the linearity 
of the amplitude evolution, which implies that 
only second powers of $c_0,c_1$ appear in the evolution of 
the density matrix.

Next, we consider a weak measurement, as the one used in \cite{Jordan2005}: 
two quantum point contacts, biased with potentials $V_A, V_B$, 
are coupled to a double quantum dot, having one excess electron. 
The electron being in the left (right) dot corresponds 
to the observable $\sigma$ having value $0$ ($1$).
When the electron is in the left dot, 
the transmission probabilities of one electron through the 
spin degenerate channel  
of the left or right QPC are $T_{A|0}$ and $T_{B|0}$, respectively, while, 
when the electron is in the right dot, 
they are $T_{A|1}:=T_{A|0}+\delta T_A>T_{A|0}$ and 
$T_{B|1}:=T_{B|0}-\delta T_B<T_{B|0}$. 
The larger the coupling of the dot to the QPC's, the 
larger the differences $T_{A|1}-T_{A|0}$ and $T_{B|0}-T_{B|1}$. 
The different values of the transmission probabilities are due to 
the coupling to the system
\[\Hat{H}_{int} = \Hat{\sigma} \left(\Hat{V}_A + \Hat{V}_B \right) \;,\]
where $\Hat{V}_a$ is a one-body operator on the detector's degrees of freedom. 
As a consequence, for $\sigma=0$ the transmission through the QPC 
is governed by a scattering matrix $\mathcal{S}^{(0)}_a$ corresponding to the 
Hamiltonian $\Hat{H}_a^{(0)}$, and, for $\sigma=1$, by $\mathcal{S}^{(1)}_a$ 
corresponding to $\Hat{H}_a^{(1)} = \Hat{H}_a^{(0)}+\Hat{V}_a$. 
If we write the scattering matrix as 
\begin{equation}
\mathcal{S} = 
\begin{pmatrix}
r&t'\\
t&r'
\end{pmatrix}
\end{equation}
we have that the transmission probabilities are $T=|t|^2$. 

Following Reference \cite{Averin2005}, 
we consider the Full Counting Statistics of the two detectors. 
We have that the probability of getting a number $n_A$ 
of electrons transmitted in the left QPC 
and $n_B$ in the right one within a time $\tau$ is 
\[P(n_A, n_B) = \sum_\sigma |c_\sigma|^2 P(n_A,n_B|\sigma),\]
where 
\[P(n_A,n_B|\sigma) = P_A(n_A|\sigma) P_B(n_B|\sigma).\]
In the shot noise limit $k_B T \ll e V_A, e V_B$, 
the probability distribution for each QPC is simply 
the binomial distribution: 
\[P_a(n|\sigma) =  \binom{N_a}{n} 
R_{a|\sigma}^{N_a-n} T_{a|\sigma}^{n}.\]
We put $R:= 1-T$ and $N_a:= 2eV_a \tau/h$. 
Then, 
for large observation time, 
the probability density $\Pi_a(I|\sigma)$ for the average current measured 
within $\tau$ is 
\[\Pi_a(I|\sigma) \simeq \frac{1}{\sqrt{2\pi S_{a|\sigma}/\tau}} 
\exp{\left\{-\frac{(I-I_{a|\sigma})^2}{2 S_{a|\sigma}/\tau}\right\}}.  
\]
$\Pi_a(I|\sigma)$ has a peak at $I_{a|\sigma} = 2 G_Q V_a T_{a|\sigma}$ 
($G_Q=e^2/h$ is the quantum of conductance and the factor of two accounts 
for spin degeneracy), taking the value 
$\Pi_a(I_{a|\sigma}|\sigma) \simeq \sqrt{\tau/2\pi S_{a|\sigma}}$
and a width $\Delta I_{a|\sigma}= \sqrt{S_{a|\sigma}/\tau}$, 
with the current noise 
$S_{a|\sigma}=2 G_Q e V_a R_{a|\sigma} T_{a|\sigma}$. 
If the observation time and the coupling with the double dot are such that 
$(S_{a|0}+S_{a|1})/\tau \ll (I_{a|0}-I_{a|1})^2$ 
the observation of current is sufficient to discriminate 
in which dot the electron is. The joint probability distribution 
for the two detectors has two peaks in the 
$I_A-I_B$ plane, at points 
$(I_{A|0}, I_{B|0})$ and $(I_{A|1}, I_{B|1})$, 
while it is negligible at $(I_{A|0}, I_{B|1})$ and $(I_{A|1}, I_{B|0})$. 
This means that it is practically impossible that the two detectors 
will give discording outcomes. Furthermore, the negligible probability of 
this disagreement depends weakly on the state of the system, and it is to 
be attributed to the imperfection of the detection.

Thus, we come to the conclusion that Quantum Mechanics, 
consistently applied to system and detectors, supports the 
ensemble interpretation of the wavefunction. 
Only experiment, however, can settle the issue. 
We discuss the theoretical implications of the possible 
experimental outcomes. 
\subsubsection{The experiments support the single-system interpretation of the 
wavefunction.} 
This would contradict Quantum Mechanics. Since the latter 
has been confirmed in many experiments, and provides 
reliable predictions, it could be retained by restraining its domain 
of validity, i.e. we would be 
forced to conclude that Quantum Mechanics is non-universal, 
in the sense that it is unable to describe the measurement 
process. In order to account for the latter, rules external to 
Quantum Mechanics have to be invoked, consisting in a generalized 
projection postulate for simultaneous measurements. 
\subsubsection{The experiments support the ensemble interpretation of the 
wavefunction.}
Then we could conclude that individual systems do possess 
objective properties, which are ascertained by measurement. 
We could also conclude that Quantum Mechanics is incomplete, 
in the sense that it does not describe these objective properties.  

In conclusion, we have proved that Quantum Mechanics is 
either complete or universal, but not both, and we have provided 
a simple experimental proposal which may settle the issue. 

\acknowledgments
I acknowledge comments and criticisms from D. V. Averin and G. Falci.

\end{document}